\documentclass[12pt,osajnl2,preprint,showpacs]{revtex4}  %% REVTeX 4.0
\usepackage{graphicx}
\begin{document}

%\twocolumn[

\title{Switching a plasma-like metamaterial via embedded resonant atoms exhibiting
electromagnetically induced transparency}

\author{Sangeeta Chakrabarti, S. Anantha Ramakrishna and Harshawardhan Wanare}

\affiliation{Department of Physics, Indian Institute of Technology Kanpur 208016, India}
%$^*$Corresponding author: sar@iitk.ac.in}

\begin{abstract} We theoretically demonstrate control of the plasma-like effective response of a
metamaterial composed of aligned metallic nanorods when the electric field of the incident radiation is parallel to
the nanorods. By embedding this metamaterial in a coherent atomic/molecular medium, for example silver nanorod arrays
submerged in sodium vapor, we can make the metamaterial transmittive in the forbidden 
frequency region below its plasma frequency. This phenomenon is enabled
by having Lorentz absorbers or other coherent processes like stimulated Raman absorption in the background medium
 which provide a large positive dielectric permittivity in the vicinity of the resonance, thereby rendering the 
effective permittivity positive. In particular, processes such as electromagnetically induced transparency 
are shown to provide additional control to switch and tune the new transmission bands. 

\end{abstract}
\maketitle
%]

Metamaterials displaying novel electromagnetic properties at
predetermined frequencies have generated considerable interest~\cite{sar_book}. The geometric structure
of the metamaterial units support localized electromagnetic resonances that is at the heart of their unique and enhanced
response. This has resulted in novel phenomena such as negative refractive index~\cite{dr_smith,sar_book}, 
electromagnetic invisibility~\cite{pendry_science}, subwavelength imaging using superlenses~\cite{pendry_prl00} etc. 
As metamaterials rely on their geometric configuration for their enhanced properties, 
they suffer from the drawback that once the structure is fabricated the properties are fixed. Futuristic
applications of these structured materials would require frequency 
agile, {\it reconfigurable} metamaterials with dynamic control~\cite{iitk_oe08}.
 
Recently, there have been a few proposals for the dynamic control of metamaterial response using various physical
processes. These include using Kerr nonlinearities~\cite{sar_prb}, photoconductivity of semiconductor
inclusions~\cite{padilla_prl}, magnetostrictive effects~\cite{lakhtakia}, liquid crystal 
inclusions~\cite{werner}, coherent control schemes~\cite{iitk_oe08} and optical gain~\cite{klar}. Parametric 
control of the metamaterial via embedded atomic/molecular media driven to coherence offers 
unique possibilities, in that, the dispersion and dissipation of the metamaterial can be 
dramatically transformed by applied electromagnetic radiation~\cite{iitk_oe08}. 
Coherent control schemes exploit the quantum mechanical response of the embedded atoms/molecules 
by carefully creating interfering pathways within the atoms. This has 
been used to show various counter-intuitive effects such as
Electromagnetically Induced Transparency (EIT)~\cite{boller}, ultra-slow light~\cite{hau} 
to superluminal light~\cite{dogariu}, enhancement of the 
index of refraction~\cite{scully}, etc. in homogenoeous atomic gases and vapours. 
We have recently shown that appropriate inclusion of 
resonant atomic/ molecular media within the metamaterial units can enable dynamic parametric control of the
metamaterial properties~\cite{iitk_oe08}. 

\begin{figure}[tbp]
\includegraphics[angle = -0, width = 0.8\columnwidth]{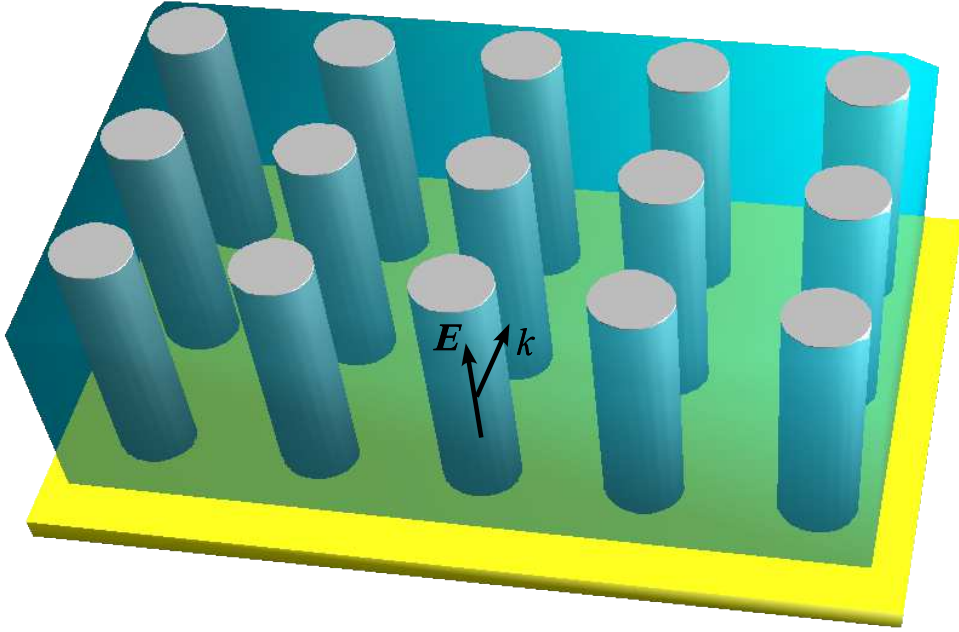}
\caption{Schematic picture of an aligned  nanorod array with subwavelength 
periodicity and grown vertically on a substrate that behaves 
as a plasma for linearly polarized radiation with the 
electric field along the wire axes. The nanorod array can be submerged in a 
background medium, for example, by placing it in a vacuum cell infused with atomic vapor, 
whose properties can then be used to switch the plasma-like behaviour.}\label{unitcell}
\end{figure}

In this letter, we consider a periodic array of parallel metallic nanorods with subwavelength
periodicity that effectively behaves like a plasma at optical and near infra-red (NIR)
frequencies for the radiation with the electric field aligned along the rods as shown in Fig.~\ref{unitcell}. 
When this metamaterial is embedded in an appropriate resonant medium, its
plasma-like response can be switched and a transmission band opens up below the cut-off (plasma) frequency.
Resonant processes such as EIT are used to tailor the properties of this transmission band. 
%For the orthogonal
%polarization with the magnetic field coaxial with the rods, one can achieve a resonant magnetic response due to the
%coupled plasmonic response of the individual rods when the rods are arranged on a closed loop within each unit cell.
%(See Fig.\ref{unitcell}.) Here, we also briefly describe dynamic control of such magnetic response.

\begin{figure}[tbp]
\includegraphics[angle = -0, width = 1.0\columnwidth]{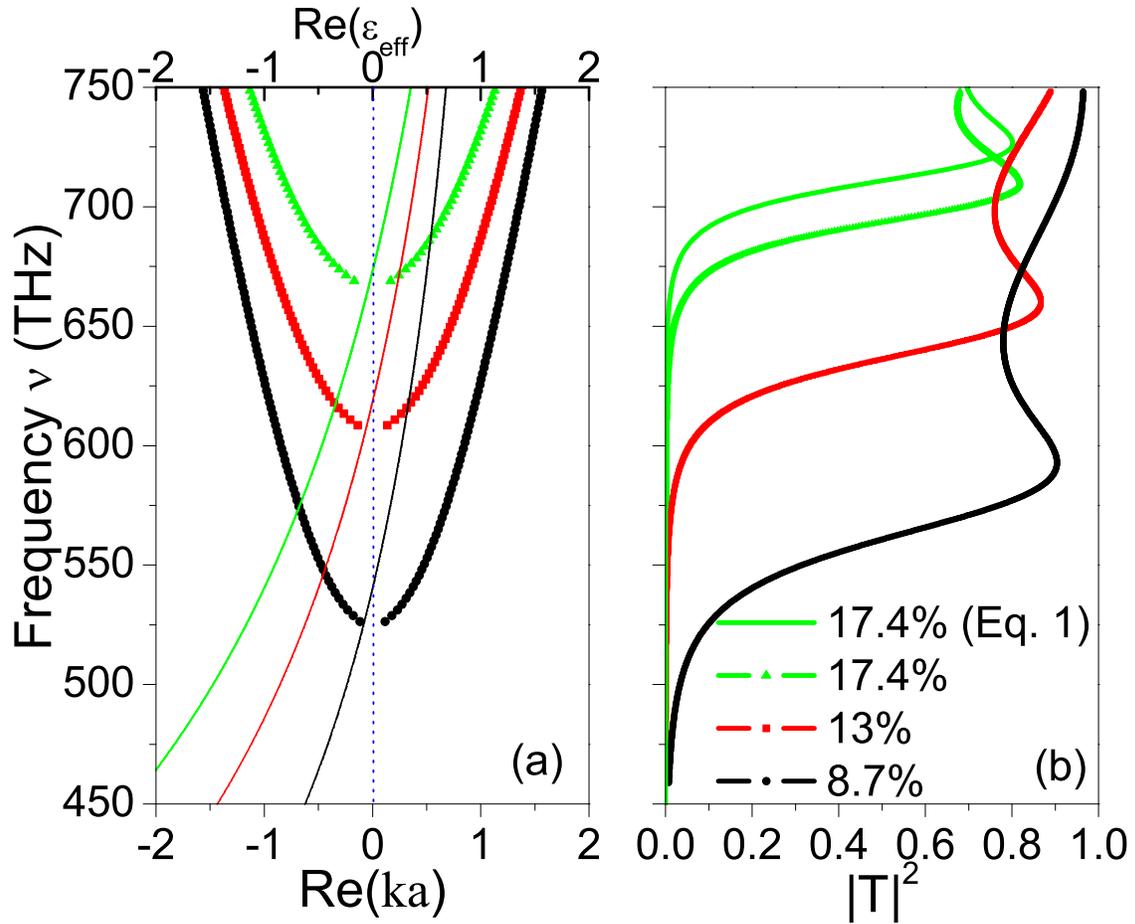}\\
\includegraphics[angle = -0, width = 1.0\columnwidth]{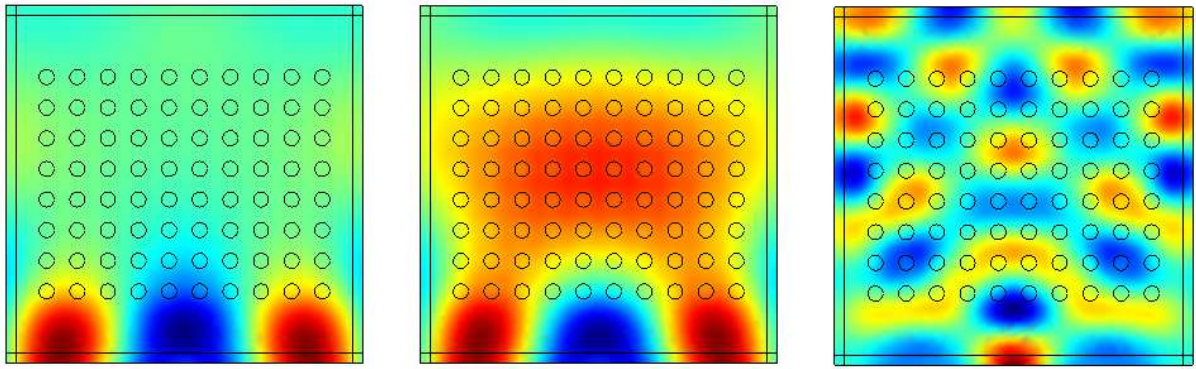}
\caption{\label{cutoff} Top panels: (a) Bandstructure of the silver nanorod metamaterial and (b) 
Transmittance from an array containing four layers of nanorods, for various 
filling fractions of silver calculated using the transfer matrix method. The polarization 
is such that the electric field is along the rod axes. The $\mathrm{Re}(\epsilon_\mathrm{eff})$ and 
transmittance calculated for a homogeneous slab of effective medium as given by 
Eq.~(\ref{effeps}) are also shown by thin lines in (a) and (b) respectively.
Bottom panels: The electric fields  (in arbitrary
units) within the rod array for a plane wave incident from the bottom. The radiation does not 
penetrate for frequencies (left, 535 THz) below the plasma frequency (middle, 537 THz) while it 
propagates across for higher frequencies (right, 540 THz).}
\end{figure}

It is well-known that the low frequency plasma-like response of sparse, conducting subwavelength thin wire 
meshes~\cite{pendry_prl96} arises due to (i) a lowered effective electron density 
($\mathrm{N_\mathrm{eff}}$) as the conducting medium is sparse, and 
(ii) an enhanced effective electronic mass ($\mathrm{m_\mathrm{eff}}$) that owes to the large inductance of
the thin wires. The resulting effective plasma frequency $\omega_p^2 = (\mathrm{N_\mathrm{eff}}e^2)/(\epsilon_0
\mathrm{m_\mathrm{eff}})$ can range from a few gigahertz to a hundred terahertz depending on the geometry. 
At optical and NIR frequencies, however, most bulk metals behave dominantly 
as plasmas rather than as Ohmic conductors. As the electromagnetic 
fields penetrate well into the metal wires, any effects arising due to the enhanced effective mass become 
less pronounced and the plasma frequency depends only on the volume fraction
occupied by the metal (filling fraction, $f$). This behaviour as a dilute metal is evidenced by the 
band structures shown in Fig.~\ref{cutoff}(a) for an array of silver nanorods. A higher metallic
filling fraction results in a larger plasma frequency ($\mathrm{Re}(\epsilon_\mathrm{eff}) > 0$ for 
$\Omega > \omega_p$ in Fig.~\ref{cutoff}(a)). 

Arrays of metal nanorods comprise one of the simplest optical metamaterials that display plasma-like behaviour. 
Large scale ordered and aligned silver nanowire arrays can be obtained, for example,  
by pulsed electrodeposition in porous alumina~\cite{sauer}.
Alternatively, vertically aligned columnar silver nanorods as schematically suggested in Fig.~\ref{unitcell},
have been grown on substrates using glancing angle e-beam deposition~\cite{zhou_jpd2008}.
The effective dielectric permittivity of the {\it dilute metal}, for radiation with 
the electric field along the nanorods, is given by 
\begin{equation} 
\epsilon_\mathrm{eff} = f \epsilon_\mathrm{m} + (1 -f)\epsilon_\mathrm{b} 
\label{effeps}
\end{equation}
 Here, the filling fraction $f = \pi r^2/a^2$, where r is the radius of the circular rods and $a$
is the periodicity of the square lattice, $\epsilon_\mathrm{m}$ and $\epsilon_\mathrm{b}$
are the dielectric permittivities of the metal and the nonconducting background medium, repectively. 
The effective plasma frequency ($\omega_p$) of the metamaterial is obtained by the condition 
$\epsilon_\mathrm{eff} = 0$. It is seen in Fig.~\ref{cutoff}(a) that propagating modes arise only 
when $\mathrm{Re}(\epsilon_\mathrm{eff}) > 0$, and in Fig.~\ref{cutoff}(b) that $\mathrm{Re}(\epsilon_\mathrm{eff})$
given by Eq.~(\ref{effeps}) reproduces the transmittance through a slab
of two layers of silver nanorods with reasonable accuracy. 
This agreement assumes significance in view of the ongoing debate on the nature and applicability of 
homogenization theories to such thin-wire metamaterials at optical frequencies~\cite{joa,bratkovsky,lakh_OC2009}
The field maps calculated by finite elements method with perfectly matched boundary layers 
confirm the plasma-like behaviour where there is very little penetration 
within the rod arrays below $\omega_p$, while the fields are transmitted across the
nanorod array above $\omega_p$ (see Fig.~\ref{cutoff}, bottom panels).

Usually $\mathrm{Re}(\epsilon_\mathrm{m})$ is a negative number with a large magnitude at optical and NIR
frequencies and a small filling fraction of the metal is sufficient to render $\epsilon_\mathrm{eff} < 0$, 
as can be deduced from Eq.~\ref{effeps}. If the
embedding medium, however, had a resonance below the effective plasma frequency, the resonant enhancement of the
$\mathrm{Re}(\epsilon_\mathrm{b})$ in the positive regime could offset the negative contribution 
of the metal. The metamaterial as a whole would now act as a positive dielectric medium and 
display a transmission band at frequencies within the width of the resonance below the resonance frequency.
Note that the transmission band occurs below the plasma frequency of the bare plasma metamaterial.
It is, indeed, counter-intuitive that introduction of a lossy resonance can induce the medium to be more transmittive. 

Resonant enhancement of the positive permittivity of the background material can be achieved by various means: 
Examples range from off-resonant Raman transitions to simple resonant absorption by the atoms / molecules
and coherent processes like EIT. Let us consider a specific and 
realistic case of silver nanorods submerged in a background of atomic sodium gas that is readily accomplished by  
placing the nanorod  array within a vacuum  cell infused with sodium vapor. 
For the frequencies of the $\mathrm{D_1}- \mathrm{D_2}$ lines of sodium, 
we construct a {\it dilute metal} (nanorod  array) with
a plasma frequency $\omega_p > \omega_{D1,~D2}$. A filling fraction of $f = 0.174$ completes the prescription for a
silver nanorod array. For example, we consider a metamaterial with a square unit cell of $a = 120$nm and
 consisting  of four square silver nanorods with sides of $b  = 25 $nm per unit cell. 
It is estimated that one can achieve a maximum of $\epsilon_\mathrm{b} \sim 2.2$ near the resonance
for pressure broadened sodium with a number density of $\sim 1 \times 10^{17}\mathrm{cm^{-3}}$~\cite{agarwal_boyd}. 
Using such a background medium, our formula predicts a maximum of
$\epsilon_\mathrm{eff} \sim 0.86$,  i.e., it leads to a positive dielectric permittivity band below $\omega_p$. 

There are alternative proposals for the creation of a transmission band below the plasma frequency, 
but for homogeneous plasmas. Harris~\cite{harris_plasma} has proposed that processes analogous to EIT 
in atoms can be used to drive a longitudinal plasma oscillation in an ideal plasma, giving rise to a 
pass band below the plasma cutoff. Agarwal and Boyd~\cite{agarwal_boyd} have proposed the elimination 
of the negative dielectric bandgap in a resonant homogeneous optical medium
by EIT effects. In contrast, we address plasma-like 
metamaterials that are inhomogeneous by construction and utilize the resonant 
enhancement of the positive dielectric permittivity of the background medium.

\begin{figure}[tbp]
\includegraphics[angle = -0, width = 0.9\columnwidth]{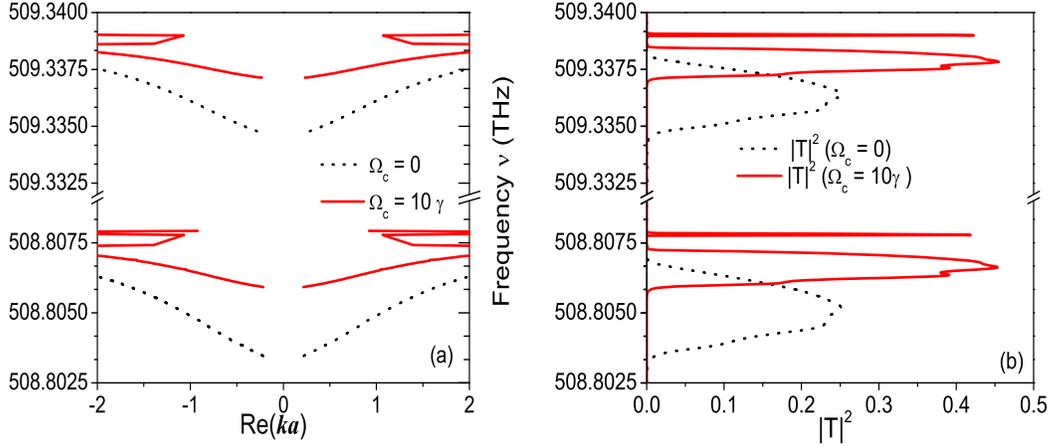}
\caption{(a) Band structure for s-polarized light with the metamaterial immersed in atomic 
sodium showing the new transmission band that develops below $\omega_p$, for the $D_1$ and $D_2$ lines. 
(b) The reflectance and transmitance corresponding to the band structure in (a). (c)}\label{splots}
\end{figure}

We now present numerical solutions for a plasma metamaterial consisting of silver nanorods immersed in a 
background vapour of sodium. We use nanorods of square cross-sections to avoid stair--casing effects on a square grid, 
while of course, maintaining the required filling fraction. The calculations are essentially two-dimensional
within the plane perpendicular to the axes of the cylinders. Invariance along the cylindrical axes
was assumed and the calculations were performed using the transfer matrix
method~\cite{tmm}. We have used experimentally determined values
for the dielectric permittivity of silver~\cite{johnson_christy} and the sodium vapour~\cite{aiaa} in our calculations.
In Fig.~\ref{splots}, new propagating bands below the effective plasma frequency ($\sim 650$~THz), 
in the presence of the resonant sodium vapour, are shown. Fig.\ref{splots}(a) shows  the band 
structure for the composite plasma metamaterial and  Fig.\ref{splots}(b) shows the transmittance 
from a slab of four layers of this metamaterial. Narrow transmission bands are found to develop at $\sim$ 508.805 
THz and 509.335 THz corresponding to the frequencies of the $D_1$ and $D_2$ lines of sodium respectively. 

Application of another control radiation 
field can dramatically transform the dielectric response and the dispersion of the sodium lines 
via coherent processes like EIT. The control field for EIT can be readily applied in the 
scheme Fig.~\ref{unitcell} that either propagates along the nanorod axis or 
co-propagates with the probe field, but with the magnetic field oriented along the
nanorod axis. Application of  a control field with a Rabi frequency $\Omega_c = 10 \gamma$ is shown to 
tune and switch the transmittive bands of the $D_1$ and $D_2$ lines in  Fig.~\ref{splots}(a)
and \ref{splots}(b). The intensity of the control field is used as a control parameter to switch the 
transmission. We find that the observed effects are reasonably independent of the exact location and 
distribution of the nanorods in the unit cell, as long as the filling fraction 
remains unaltered.  We also note that  
plasma-like metallic metamaterials with even larger filling fractions can be also 
switched using a coherently enhanced refractive index for the background atomic gas~\cite{scully}.
Finally and crucially, we also would point out that the present proposal for switching the plasma metamaterial
depends on making the effective permittivity positive through a bulk volume averaging process. This is 
very different from our previous work~\cite{iitk_oe08} on parametrically transforming 
the resonances of the metamaterial units via coherent processes. 

In conclusion, we demonstrate  the posssibility of introducing a propagating band below the 
plasma frequency by coherent optical processes in a composite metallic nanorod array with a resonant atomic
background. The resonant response of the positive permittivity 
of the background medium can be sufficient to cancel out the negative permittivity of the dilute plasma leading to a net
positive effective medium permittivity. Coherent optical processes such as EIT are
 used to further engineer this new transmission band.


\begin{thebibliography}{99}
\bibitem{sar_book} S. A. Ramakrishna and T.M. Grzegorczyk, {\it Physics and applications of negative refractive 
index materials} (CRC Press, Boca Raton USA, 2009)
%\bibitem{veselago} V. G. Veselago, {\it Electrodynamics of substances with simultaneously negative values of $\epsilon$ and $\mu$}, Sov. Phys. Usp. {\bf 10}, 509 (1968).
\bibitem{dr_smith} D. R. Smith, Willie J. Padilla, D. C. Vier, S. C. Nemat-Nasser, and S. Schultz, {\it Composite medium with simultaneously negative permeability and permittivity}, Phys. Rev. Lett., {\bf 84}, 4184 (2000)
\bibitem{pendry_science} J.B. Pendry, D.R. Smith and D. Schurig, {\it Controlling electromagnetic fields}, Science, {\bf 312}, 1780 (2006)
\bibitem{pendry_prl00} J.B. Pendry, {\it Negative refraction makes a lerfect lens}, Phys. Rev. Lett., {\bf 85}, 3966 (2000)
\bibitem{sar_prb}S. O'Brien, D. McPeake, S. A. Ramakrishna, and J. B. Pendry, {\it Near infrared photonic bandgaps and nonlinear effects in negative magnetic metamaterials}, Phys. Rev. B, {\bf 69}, 241101 (2004) 
\bibitem{padilla_prl} W. J. Padilla, A. J. Taylor, C. Highstrete, Mark Lee, and R. D. Averitt, {\it Dynamical electric and magnetic metamaterial response at terahertz frequencies}, Phys. Rev. Lett., {\bf 96}, 107401 (2006)
\bibitem{lakhtakia} J. Han, A. Lakhtakia and C.-W. Qiu, {\it Terahertz metamaterials with semiconductor split-ring resonators for magnetostatic tunability}, Optics Express, {\bf 16}, 14390 (2008)
\bibitem{werner}D.H. Werner, D.-H. Kwon, I.-C. Khoo A.V. Kildishev and V.M. Shalaev, {\it Liquid crystal clad near-infrared metamaterials with tunable negative-zero-positive refractive indices}, Opt. Express {\bf 15}, 3342 (2007).
\bibitem{iitk_oe08} S. Chakrabarti, S. A. Ramakrishna and H. Wanare, {\it Coherently controlling metamaterials}, Opt. Express, {\bf 16}, 19504 (2008)
\bibitem{klar} T.A. Klar, A.V. Kildishev, V.P. Drachev, and V.M. Shalaev, {\it Negative-index metamaterials: Going optical}, 
IEEE J Selected Topics Quant. Electr. {\bf 12}, 1106 (2006).
\bibitem{boller} K. J. Boller, A. Imamoglu, A. \& S. E. Harris, {\it Observation of electromagnetically induced transparency}, Phys. Rev. Lett., {\bf 66}, 2593 (1991)
\bibitem{hau} C. Liu, Z. Dutton, C. H. Behroozi \& L. V. Hau, {\it Light speed reduction to 17 meters per second in an ultracold atomic gas}, Nature, {\bf 397}, 594 (1999)
\bibitem{dogariu} L. J. Wang, A. Kuzmich, \& A. Dogariu, {\it Gain-assisted superluminal light propagation}, Nature, {\bf 406}, 277 (2000)
%\bibitem{dutton} C. Liu, Z. Dutton, C. H. Behroozi \& L. V. Hau, {\it Observation of coherent optical information storage in an atomic medium using halted light pulses}, Nature, {\bf 409}, 490 (2001)
\bibitem{scully} M.O. Scully, {\it Enhancement of the index of refraction via quantum coherence}, Phys. Rev. Lett., {\bf 67}, 1855 (1991)
\bibitem{pendry_prl96} J. B. Pendry, A. J. Holden, W. J. Stewart, and I. Youngs, {\it Extremely low frequency plasmons in metallic mesostructures}, Phys. Rev. Lett., {\bf 76}, 4773 (1996)
\bibitem{sauer} G. Sauer, G. Brehm, S. Schneider, K. Nielsch, R. B. Wehrspohn, J. Choi, H. Hofmeister and U. G\"{o}sele, {\it Highly ordered monocrystalline silver nanowire arrays}, J. Appl. Phys., {\bf 91}, 3243 (2002)
\bibitem{zhou_jpd2008} Q. Zhou, Z. Li, Y. Yang, Z. Zhang, {\it Arrays of aligned, single crystalline silver nanorods for 
trace amount detection}, J. Phys. D: Appl. Phys. {\bf 41}, 152007 (2008).
\bibitem{joa} A.I. Rahachou and I.V. Zozoulenko, {\it Light propagation in nanorod arrays}, 
J. Opt. A: Pure Appl. Opt. {\bf 9}, 265 (2007)
\bibitem{bratkovsky}A. Bratkovsky, E. Ponizovskaya, S.-Y. Wang, P. Holmstrom, L. Thylen, Y. Fu, and H. Agren {\it A metal-wire /quantum-dot composite metamaterial with negative $\epsilon$ and compensated optical loss}, Appl. Phys. Lett. {\bf 93}, 
193106 (2008).
\bibitem{lakh_OC2009} T.G. Mackay and A. Lakhtakia, {\it On the application of homogenization formalisms to active dielectric composite metamaterials}, Opt. Commun. {\bf 282}, 2470 (2009).
%\bibitem{pendry_jpcm98} J. B. Pendry, A. J. Holden, D. J. Robbins and W. J. Stewart, {\it Low frequency plasmons in thin wire structures},J. Phys.: Condens. Matt., {\bf 10}, 4785 (1998)
\bibitem{harris_plasma} S. E. Harris, {\it Electromagnetically induced transparency in an ideal plasma}, Phys. Rev. Lett., {\bf 77}, 5357 (1996)
\bibitem{agarwal_boyd}G. S. Agarwal and Robert W. Boyd, {\it Elimination of the band gap of a resonant optical material by electromagnetically induced transparency}, Phys. Rev. A. {\bf 60}, R2681 (1999)
%\bibitem{raman} 
\bibitem{tmm} J.B. Pendry and A. Mackinnon, {\it Calculation of photon dispersion relations}, Phys. Rev. Lett. {\bf 69}, 2772 (1992)
\bibitem{johnson_christy} P.B. Johnson and R.W. Christy, {\it Optical constants of the noble metals}, Phys. Rev. B, {\bf 6},  4370  (1972)
\bibitem{aiaa} G. Blendstrup, D. Bershader and P. W. Langhoff, {\it Resonance refractivity studies of sodium vapor for enhanced flow visualization}, AIAA Journal, {\bf 16}, 1106 (1978)
\end{thebibliography}
\end{document}